\documentclass[a4paper]{article}

\usepackage[utf8]{inputenc}
\usepackage{amsmath}
\usepackage{amsfonts}
\usepackage{amssymb}
\usepackage{amsthm}
\usepackage{enumerate}
\usepackage{color}
\usepackage{url}
\usepackage[english]{babel}
\usepackage{microtype}

\title{Reed--Solomon--Gabidulin Codes}

\author{%
Xavier Caruso\footnote{CNRS, Institut Mathématique de Bordeaux, équipe LFANT}~
and Amaury Durand\footnote{Université de Franche-Comté}}

\newtheorem{theorem}{Theorem}[section]
\newtheorem{lemma}[theorem]{Lemma}
\newtheorem{proposition}[theorem]{Proposition}

\theoremstyle{definition}
\newtheorem{definition}[theorem]{Definition}
\newtheorem{remark}[theorem]{Remark}
\theoremstyle{remark}
\newtheorem{example}[theorem]{Example}

\newcommand{\QQ}{\mathbb Q}
\newcommand{\FF}{\mathbb F}

\newcommand{\id}{\textrm{\rm id}}
\newcommand{\End}{\textrm{\rm End}}

\newcommand{\Frob}{\textrm{\rm Frob}}
\newcommand{\ev}[1]{\textrm{\rm ev}_{#1}}
\renewcommand{\mod}{\,\%\,}

\newcommand{\rgcd}{\textsc{rgcd}}
\newcommand{\llcm}{\textsc{llcm}}

\newcommand{\bc}{\textbf{c}}
\newcommand{\bg}{\textbf{g}}
\newcommand{\RSG}{\textrm{\rm RSG}}
\newcommand{\wH}{w_{\textrm{\rm H}}}
\newcommand{\wrH}{w_{\textrm{\rm rH}}}
\newcommand{\drH}{d_{\textrm{\rm rH}}}
\newcommand{\good}{\textrm{\rm good}}

\begin{document}
\maketitle

\begin{abstract}
We introduce Reed-Solomon-Gabidulin codes which is, at the same time, an 
extension to Reed-Solomon codes on the one hand and Gabidulin codes on 
the other hand. We prove that our codes have good properties with 
respect to the minimal distance and design an efficient decoding 
algorithm.
\end{abstract}

\subsection*{Important disclaimer}

After we made this article available on HAL and arXiv, we received an 
email from Umberto Martinez-Penas, in which he kindly explained to us 
that the results obtained in the present article were already discovered 
(and partly published) recently~\cite{martinez1, martinez2}; our notion 
of Reed--Solomon--Gabidulin codes is actually a special case of the 
notion of Linearized Reed-Solomon codes introduced there.

Nevertheless, our exposition differs a bit from that of \emph{loc. cit}, 
so we think that our article still has some interest. Combined with the 
results of~\cite{carleb}, our version of the decoding algorithm has 
sub-quadratic complexity; this was left as an open question 
in~\cite{martinez2}.

\section*{Introduction}

Reed--Solomon codes form a well-known class of error detection and 
correction codes which have very interesting properties (optimal minimal 
distance, efficient decoding algorithms). They were introduced in 1960 
by Reed and Solomon and are nowadays widely used in everyday life. About 
twenty years later, Delsarte~\cite{delsarte}, Gabidulin~\cite{gabidulin} 
and Roth~\cite{roth}---independently---imagined an analogue of 
Reed--Solomon codes in the context of the rank distance, which is finer 
than the standard Hamming distance and well suited for some applications 
(\emph{e.g.} network coding). These codes are nowadays called 
\emph{Gabidulin codes}. Their construction is based on the concept of 
linearized polynomials over the finite fields. More recently several 
authors generalized and optimized Gabidulin codes. In 2013, in her 
thesis~\cite{wachter} and subsequent papers, Wachter-Zeh proposed an 
efficient implementation of operations with linearized polynomials, 
together with an equivalent of Gao's decoding algorithm.

In 2009, Boucher, Geiselmann and Ulmer~\cite{bougeiulm} introduced 
analogues of BCH codes in the Gabidulin's context of linearized 
polynomials (\emph{cf} also~\cite{bouulm}). 
It worths mentionning that they use Ore 
polynomials (introduced by Ore in 1933 in~\cite{ore}) in place of 
linearized polynomials. Although the two approaches are equivalent in 
the case of finite fields, it turns out that Ore polynomials are more 
general objects which continue to make sense in a large variety of 
settings.
Taking advantage of this new point of view, Robert proposed in his
thesis~\cite{robert} an extension of Gabidulin's code to the
caracteristic zero, in which basically finite fields are replaced 
by number fields.

Another advantage of Boucher, Geiselmann and Ulmer's approach is that it 
allows longer codes: while the length of a Gabidulin code is necessarily 
bounded from above by the degree of the finite field we are working 
with, this bound can be generally overpassed in Boucher, Geiselmann and 
Ulmer's construction. On the other hand, no efficient decoding algorithm 
is known.

\paragraph{Contribution of the article.}

In the present paper, we introduce and study a new generalization 
of Gabidulin codes, which combines all the benefits of previous 
constructions. Precisely, we shall show that:

\noindent
(1) as for Gabidulin codes, our codes are MDS (Maximal Distance
Separable),

\noindent
(2) as in Boucher, Geiselmann and Ulmer's work, long codes are 
permitted,

\noindent
(3) as in Wachter-Zeh's work, there exists an efficient decoding 
algorithm.

\smallskip

\noindent 
Besides, the setting we consider includes the case of finite fields (as 
in Gabidulin's initial definition) and number fields (as in Robert's 
generalization) but it is even more general. 
For example, our construction allows the base field to be the field of 
rational fractions in the variable $t$ over a finite field equipped with 
its canonical derivation~$\frac d{dt}$.

Moreover it turns out that, for a special choice of parameters, our 
codes extend classical Reed--Solomon codes. For this reason, we have 
decided to call them \emph{Reed--Solomon--Gabidulin (RSG\footnote{Be 
careful at not making the confusion with GRS codes, which stands for 
\emph{Generalized Reed--Solomon codes}.} for short) codes}.

\paragraph{Organization of the article.}

This paper is divided in two sections. 
The first one is devoted to introduce and develop the necessary 
background on Ore polynomials and related notions. We will study
particularly the notion of \emph{evaluation morphisms} which is 
the main ingredient we will need for defining GRS codes.
In the second section, we introduce GRS codes and state their main 
properties (\emph{cf} (1), (2), (3) above). For the sake of brievity, 
proofs are omitted though intermediate steps are often isolated.

\section{Ore polynomials}

Throughout this article, we use the following notation: $K$ is a field, 
$\theta : K \to K$ be a ring homomorphism and $\partial : K \to K$ be a 
$\theta$-derivation, i.e. an additive mapping such that $\partial(ab) = 
\theta(a)\partial(b) + \partial(a)b$ for all $a,b \in K$.

We shall denote by $F$ the subfield of $K$ consisting of elements
$a$ such that $\theta(a) = a$ and $\partial(a) = 0$. 
\textbf{We will always assume that the extension $K/F$ is finite}
and will denote by $r$ its degree. Our assumption implies in particular 
that $\theta$ has finite order and thus is bijective.

\begin{definition}[Ore polynomial ring]
The ring of Ore polynomials $K[X; \theta, \partial]$ is the ring 
whose elements are polynomials in $X$ over $A$ endowed with the usual 
addition and with the multiplication defined by the rule:
$$X \times a = \theta(a)X + \partial(a), \quad \forall a \in A.$$
\end{definition}

\begin{example}
\label{ex:Ore}
Throughout this article, we will illustrate our constructions 
with the two following examples:
\begin{itemize}
\renewcommand{\itemsep}{1ex}
\item[(1)] (This setting is the one in which Gabidulin codes were
first defined by Gabidulin in~\cite{gabidulin}, with a slightly different
vocabulary.)
Let $p$ be a prime number, $q$ be a power of $p$ and $r$
be a positive integer. We let $\FF_{q^r}$ denote a finite field
with cardinality $q^m$. We endow it with the Frobenius $\Frob_q :
x \mapsto x^q$. The first Ore ring we will be interested in is
$\FF_{q^r}[X ; \Frob_q, 0]$. In this setting, the subfield $F$ of 
$K = \FF_{q^r}$ we have introduced is $\FF_q$. The degree of the
extension $K/F$ is then $r$.
\item[(1')] More generally, one can pick an arbitrary field $K$,
endow it with a finite order automorphism $\theta$ and consider the
Ore ring $K[X,\theta,0]$. Beyond the case of finite fields, natural
examples are cyclotomic extensions of $\QQ$ or Kummer extensions.
This case was addressed in Robert's thesis~\cite{robert}.
\item[(2)] Let $\kappa$ be a field of characteristic $p$. We consider 
the field $K = \kappa(t)$ and endow it with the natural derivation 
$\frac d{dt}$. We can then form the Ore ring $\kappa(t)[X,\id,\frac 
d{dt}]$. Here the subfield $F$ of $K$ is $\kappa(t^p)$ and the degree 
of the extension $K/F$ is then $p$.
\end{itemize}
\end{example}

The notion of degree extends \emph{verbatim} to Ore polynomials: if $P = 
\sum a_iX^i$ is an Ore polynomial, its degree is the largest integer $i$ 
for which $a_i \neq 0$.
Besides, one can prove the existence of a right Euclidean division for 
Ore polynomials: if $A, B \in K[X;\theta,\partial]$ with $B \neq 0$, 
there exist unique $Q, R \in K[X;\theta,\partial]$ with $A = QB+R$ and 
$\deg R < \deg B$. This has the usual consequences: the noncommutative
ring $K[X;\theta,\partial]$ is left-principal, right \textsc{gcd}s and
left \textsc{lcm}s are well defined and can be computed by Euclidean
algorithm. 
Similarly, left Euclidean divisions, left \textsc{gcd}s and right 
\textsc{lcm}s do exist (since our general assumptions imply that
$\theta$ is bijective).

\medskip

\noindent
\textit{Notation:}
In what follows, we will denote by $A \mod B$ the remainder in the 
right division of $A$ by $B$.

\subsubsection*{The centre.}

Recall that the centre of a noncommutative ring $A$ is by definition
the subset of $A$ consisting of elements $x$ such that $xy = yx$ for
all $y \in A$. We observe in particular that the centre of $A$ is a
commutative subring of $A$.
In the case of Ore polynomials, the centre can actually be computed
precisely.
In what follows, we will not need a complete description but only
the general structure of the centre as given by the next proposition.

\begin{proposition}
\label{prop:centre}
There exists a central Ore polynomial $Z(X) \in K[X; \theta, \partial]$ of 
degree $r$ such that the centre of $K[X; \theta, \partial]$ is 
$F[Z(X)]$, \emph{i.e.} the subset of Ore polynomials that can be
written as a polynomial in $Z(X)$ with coefficient in $F$.
\end{proposition}

We observe that the equality:
$$a_0 + a_1 Z(X) + \cdots + a_d Z(X)^d 
= b_0 + b_1 Z(X) + \cdots + a_e Z(X)^e$$
implies readily that $d = e$ (compare the degrees) and $a_i = b_i$ 
for all $i$. As a consequence the centre $F[Z(X)]$ is an actual
(commutative) ring of univariate polynomials with coefficients in $F$.

On the other hand, we draw the attention of the reader to the fact that 
the properties of Proposition~\ref{prop:centre} do not determine $Z(X)$ 
uniquely but only up to an additive constant in $F$.

\begin{example}
\label{ex:centre}
We continue Example \ref{ex:Ore}. In the settings~(1) and~(1'), it is 
easily seen that the centre of $K[X;\theta,0]$ is $F[X^r]$.
In the setting~(2), the centre of $\kappa(t)[X; \id, \frac d{dt}]$ (where
$\kappa$ is a field of characteristic $p$) is $\kappa(t^p)[X^p]$. 
\end{example}

\subsubsection*{Pseudo-linear morphisms.}

Another important notion is that of pseudo-linear morphisms. It is 
defined as follows:

\begin{definition}[Pseudo-linear morphism]
Let $M$ and $N$ be two vector spaces over $K$.
A \emph{pseudo-linear morphism} $u : M\to N$ is a map verifying 
$u(ax) = \theta(a)u(x) + \partial(a)x$ for all $a \in K$ and $x \in M$.
\end{definition}

We observe that any pseudo-linear morphism is \emph{a fortiori}
$F$-linear (where $F$ is defined at the beginning of this section).

Pseudo-linear morphisms are relevant in the context of Ore polynomials 
because the Ore multiplication reflects the composition rule of 
pseudo-linear morphisms. More precisely, given a pseudo-linear 
endomorphism $u : M \to M$ and an Ore polynomial $P = \sum_i a_i X^i \in 
K[X;\theta,\partial]$, one defines $P(u) = \sum_i a_i u^i$. One then 
easily checks that $P(u) \circ Q(u) = (PQ)(u)$ where the multiplication 
on the right hand size is the Ore multiplication. In other words, 
denoting by $\End_F(M)$ the ring of $F$-linear maps from $M$ to itself, 
the ``evaluation'' mapping
$$\ev{u} : \quad K[X;\theta,\partial] \to \End_F(M), \quad
P(X) \mapsto P(u)$$
is a ring homomorphism for any pseudo-linear endomorphism $u$.

The case where $M$ is $K$ itself deserves particular attention.
Indeed, we first observe that evaluation is then closely related to
Euclidean division thanks to the formula:
\begin{equation}
\label{eq:evanddiv}
\textstyle \ev{u}(P)(a) = 
a \cdot P \mod \big(X - \frac{u(a)}{a}\big)
\end{equation}
which is correct for any pseudo-linear endomorphism $u$ of $K$, any $P 
\in K[X;\theta, \partial]$ and any $a \in K$. Second, we have a complete 
classification of pseudo-linear endomorphisms of $K$.

\begin{proposition}
The pseudo-linear endomorphisms of $K$ are exactly the maps of
the form $\partial + c\theta$ with $c \in K$.
\end{proposition}

In what follows, we will often use the notation $\ev c$ in place of 
$\ev{\partial + c \theta}$.

\paragraph{Main properties of the $\ev c$'s.}

We denote by $K_\good$ the subset of $K$ consisting of elements $c$ for 
which $\partial + c\theta$ is not of the form $a{\cdot}\id$ with $a \in 
F$. Except in the very particular case where $\theta = \id$ and 
$\partial = 0$ (where $K_\good$ is obviously empty), one can prove that 
there is at most one bad value of~$c$, \emph{i.e.} the difference 
between $K$ and $K_\good$ consists at most of one element.

\begin{proposition}
\label{prop:evc}
For all $c \in K_\good$, the ring homomorphism $\ev{c}$ is surjective
and its kernel is a principal ideal generated by $Z(X) - N(c)$
for some element $N(c) \in F$.
\end{proposition}

\begin{remark}
The function $N$ defined by Proposition \ref{prop:evc} above is not 
canonical since it depends on the choice of the constant coefficient 
of $Z(X)$. Two different choices lead to functions $N$ and $N'$ such
that $N' = N + a$ for some constant $a \in F$.
\end{remark}

\begin{definition}
\label{def:equiv}
Let $c_1, c_2 \in K_\good$.
We say that $c_1$ and $c_2$ are \emph{equivalent} if
$\ker \ev{c_1} = \ker \ev{c_2}$ or, equivalently, $N(c_1) = N(c_2)$.
\end{definition}

\noindent
Using Noether--Skolem Theorem, one can prove the following
characterization:

\begin{lemma}
\label{lem:equiv}
The elements $c_1$ and $c_2$ are equivalent if and only if there exists 
$a \in K$, $a \neq 0$ such that $c_1 a = c_2 \theta(a) + \partial(a)$.

\noindent
In particular, the equivalence class of $c \in K$ is exactly the image 
of $x \mapsto \frac{(\partial + c\theta)(x)} x$.
\end{lemma}

\begin{example}
\label{ex:equiv}
Let us first focus on the settings~(1) and~(1') of Example~\ref{ex:Ore}. 
The subset $K_\good$ is then $K \backslash \{0\}$. Moreover if we have 
chosen $Z(X) = X^r$ (see Example~\ref{ex:centre}), it is not difficult
to prove that the map $N$ is the norm of $K$ over $F$.
In this context, the characterization of Lemma~\ref{lem:equiv} is 
a classical consequence of Hilbert 90 theorem which says that an
element has norm $1$ if and only if it can be written 
$\frac{\theta(a)} a$ for some $a \neq 0$.

\noindent
When $K = \FF_{q^m}$ and $\theta = \Frob_q$, we have 
$N(c) = c^{1 + q + q^2 + \cdots + q^{m-1}}$. In this case, the image of 
$N$ is $\FF_q^\star$ and there is exactly $q{-}1$ equivalence classes 
for the equivalence relation introduced in Definition~\ref{def:equiv}.

In the setting~(2), we have $K_\good = K$. Moreover, with the 
normalization $Z(X) = X^p$, one can prove\footnote{Through the proof is 
not obvious.} that $N(f) = \frac {d^{p-1}f}{dt^{p-1}} + f^p$ for any $f 
\in k(t)$. Here, Lemma~\ref{lem:equiv} asserts that $N(f) = N(g)$ if
and only if the difference $f-g$ is a logarithmic derivative.
It is easily seen that a polynomial cannot be a logarithmic derivative.
Consequently the elements of $\kappa[t]$ are pairwise nonequivalent,
implying in particular that there are infinitely many equivalence
classes for this relation.
\end{example}

\section{Reed--Solomon--Gabidulin codes}

We keep the notations of the previous section. In particular, we recall 
that $K_\good$ is the subset of $K$ consisting of elements $c$ for which 
$\partial + c\theta$ is not of the form $a{\cdot}\id$ with $a \in F$.

\subsubsection*{Setting.}

Throughout this section, we fix a positive integer $s$. We consider a 
family $\bc = (c_1, \ldots, c_s)$ of $s$ elements of $K_\good$ which are 
pairwise non-equivalent in the sense of Definition \ref{def:equiv}.
Moreover, for each $i \in \{1,\ldots,s\}$, we pick a positive integer
$n_i$ together with a family $\bg_i = (g_{i,1}, \ldots, g_{i,n_i})$ of 
$F$-linearly independant elements of $K$. The latter condition obviously
implies that $n_i \leq [K:F]$ for all $i$.
We set $n = n_1 + \ldots + n_s$.
To all these data, we associate the $K$-linear mapping:
$$\begin{array}{rcl}
\gamma_{\bc,\bg} : \, K[X;\theta,\partial] & \longrightarrow 
 & K^{n_1} \times K^{n_2} \times \cdots \times K^{n_s} \smallskip \\
P(X) & \mapsto 
 & \big(\ev{c_1}(P)(g_{1,1}), \ev{c_1}(P)(g_{1,2}), \ldots, \ev{c_1}(P)(g_{1,n_1}), \\
&& \phantom{\big(}\ev{c_2}(P)(g_{2,1}), \ev{c_2}(P)(g_{2,2}), \ldots, \ev{c_2}(P)(g_{2,n_2}), \\
&& \phantom{\big(}\ldots, \\
&& \phantom{\big(}\ev{c_s}(P)(g_{s,1}), \ev{c_s}(P)(g_{s,2}), \ldots, \ev{c_s}(P)(g_{s,n_s})\big)
\end{array}$$
Thanks to Eq.~\eqref{eq:evanddiv}, the mapping $\gamma_{\bc,\bg}$
can be rewritten in terms of Euclidean divisions. More precisely,
for $1 \leq i \leq s$ and $1 \leq j \leq n_i$, letting:
\begin{equation}
\label{eq:aij}
a_{i,j} = \frac{(\partial + c_i\theta)(g_{i,j})}{g_{i,j}}
\end{equation}
we have $\ev{c_i}(g_{i,j}) = g_{i,j} \cdot P \mod (X - a_{i,j})$.

For any positive $k$, we let $\gamma_{k,\bc,\bg}$ denote the 
restriction of $\gamma_{\bc,\bg}$ to the subspace 
$K[X;\theta,\partial]_{<k}$ consisting of Ore polynomials of
degree less than $k$.

\begin{example}
\label{ex:aij1}
Consider the setting~(1) of Example~\ref{ex:Ore}. Let $g$ be a 
multiplicative generator of $\FF_{q^r}^\star$. Its norm over $\FF_q$ is 
a multiplicative generator of $\FF_q^\star$. By what we did in 
Example~\ref{ex:equiv}, the elements $c_i = g^i$ for $0 \leq i < s$ 
are pairwise nonequivalent as soon as $s \leq q-1$. (Here, for
simplicity, we have shifted our indices so that they start from $0$
instead of $1$.)
Moreover $(1, g, \ldots, g^{r-1})$ is a 
basis of $\FF_{q^r}$ over $\FF_q$. One can then take $n_i = r$ for all 
$i$ and $g_{i,j} = g^j$ for $0 \leq j < r$. With these parameters, we 
easily compute $a_{i,j} = c_i \cdot \Frob_q(g_{i,j}) \cdot g_{i,j}^{-1} 
= g^{i + (q-1)j}$.
\end{example}

\begin{example}
\label{ex:aij2}
Consider the setting~(2) of Example~\ref{ex:Ore}.
By Example~\ref{ex:equiv} again, we can take any family $(c_1, \ldots, 
c_s)$ of pairwise distinct polynomials. Moreover a basis of $\kappa(t^p)$ 
over $\kappa(t)$ is obviously $(1, t, \ldots, t^{p-1})$. Therefore, we can 
take $n_i = p$ and $g_{i,j} = t^j$ for $0 \leq j < p$. A direct 
computation leads to $a_{i,j} = \frac j t + c_i$. 

\noindent
Taking $\kappa = \FF_3$, $k = 5$, $\bc = (0,1)$ and $\bg = 
((1,t,t^2),(1,t,t^2))$, we find that the matrix of
$\gamma_{k,\bc,\bg}$ is:
\begin{equation}
\label{eq:gamma}
\left(\begin{array}{c@{\hspace{3ex}}c@{\hspace{3ex}}c@{\hspace{1.5ex}}|
@{\hspace{1.5ex}}c@{\hspace{3ex}}c@{\hspace{3ex}}c}
1 & t & t^2 &
1 & t & t^2 \\
0 & 1 & 2t &
1 & t{+}1 & t^2{+}2t 
\end{array} \right).
\end{equation}
\end{example}

The kernel of $\gamma_{k,\bc,\bg}$ is the principal ideal generated
by the Ore polynomial:
\begin{equation}
\label{eq:defL}
L = \llcm((X - a_{i,j})_{1 \leq i \leq m,\,1 \leq j \leq n_i}).
\end{equation}
The next lemma shows that the assumption we made on the $c_i$'s
and $g_{i,j}$'s are directly related to the degree of $L$.

\begin{lemma}
\label{lem:llcmaij}
With the above notations and assumptions, the Ore polynomial $L$
has degree $n$.

\noindent
In particular, the map $\gamma_{n,\bc,\bg}$ is bijective.
\end{lemma}

\begin{example}
Continuing Example~\ref{ex:aij1}, 
the Ore polynomial $L$ defined in \eqref{eq:defL} is
$L = \prod_{i=1}^s (X^r - N(c_i))$
where we recall that $N : \FF_{q^r} \to \FF_q$ is the norm map.
(Observe that the factors $X^r - N(c_i)$ all lie in the centre of
$\FF_{q^r}[X;\Frob_q,0]$ so that the product we have written in
not ambiguous.)
In particular, when $s = q-1$, we get $L(X) = X^{r(q-1)} - 1$.
\end{example}

\begin{example}
\label{ex:L}
Continuing Example~\ref{ex:aij2} and assuming further that the
$c_i$'s lie in $\kappa$, we find that the polynomial $L$ defined in
\eqref{eq:defL} is $L = \prod_{i=1}^s (X^p - c_i^p)$.
In particular, if $\kappa$ is a finite field of cardinality $q$ and
the $c_i$'s enumerate the elements of $\kappa$ (so that $s=q$), we
have $L(X) = X^{pq} - X^p$.
\end{example}

\subsubsection*{Definition and first properties.}

We are now ready to define Gabidulin codes in the extended framework
discussed in the introduction of this section.

\begin{definition}
With the previous notations, the \emph{Reed--Solomon--Gabidulin (RSG for 
short) code} $\RSG_{k,\bc,\bg}$ associated to $\bc$ and $\bg$ is the 
image of $\gamma_{k,\bc,\bg}$.
\end{definition}

\begin{remark}
From the definition, it follows that the matrix of $\gamma_{k,\bc,\bg}$
(in the canonical basis) is a generator matrix of $\RSG_{k,\bc,\bg}$.
The matrix \eqref{eq:gamma} then provide an example of a generator 
matrix of a RSG code.
\end{remark}

It is well known that the relevant distance for Gabidulin codes is not 
the Hamming distance but the rank distance. In the context of Gabidulin 
codes introduced above, we shall need another distance which is a 
mixture between Hamming and rank distance. It is defined as follows.

\begin{definition}
Let $x = (x_{i,j})_{1 \leq i \leq m, \; 1 \leq j \leq n_i} \in
K^{n_1} \times K^{n_2} \times \cdots \times K^{n_s}$.
The \emph{rank-Hamming weight} of $x$ is:
$$\wrH(x) = 
\sum_{i=1}^s \dim_F \big< x_{i,1}, x_{i,2}, \ldots, x_{i,{n_i}} \big>_F.$$
Given $x, y \in K^{n_1} \times K^{n_2} \times \cdots \times K^{n_s}$, 
the {rank-Hamming distance} between $x$ and $y$ is $\drH(x,y) = 
\wrH(x-y)$.
\end{definition}

\begin{remark}
The weight $\wrH$ is finer that the usual Hamming weight in the 
sense that, for all $x \in K^{n_1} \times \cdots \times K^{n_s}$, 
we have $\wrH(x) \leq \wH(x)$ if $\wH$ denotes the Hamming weight.
\end{remark}

The RSG codes we have defined extend the classical notion of Gabidulin 
codes introduced in~\cite{gabidulin}. More precisely, the latter correspond to 
the case where $s = 1$, $\partial = 0$ and $K$ is a finite field. 
Relaxing the assumption on $K$, we obtain the generalized Gabidulin 
codes defined by Robert in his thesis~\cite{robert}. In particular, in this 
case, the rank-Hamming distance is the usual rank distance.

On the other hand, when $\theta = \id$ and $\partial = 0$ (that is $F = 
K$), the notion of RSG code is nothing but the standard notion of 
Reed--Solomon code and the rank-Hamming distance reduces to the 
usual Hamming distance.

\begin{proposition}
\label{prop:mindist}
The code $\RSG_{k,\bc,\bg}$ has length $n$,
dimension $k$ and minimal distance $d = n - k + 1$.
\end{proposition}

\begin{example}
The RSG code corresponding to the generator matrix \eqref{eq:gamma}
has length $6$, dimension $2$ and minimal distance $6-2+1 = 5$. It
then corrects any error of rank-Hamming weight at most $2$.
\end{example}

\subsubsection*{Decoding Reed--Solomon--Gabidulin codes.}

RSG codes can be decoded by a noncommutative extension of Gao's 
algorithm~\cite{gao}. This fact was already observed in the works of 
Wachter-Zeh and al.~\cite{wachter} in the special case of usual Gabidulin
codes. After what we have done previously, the extension to RSG codes 
is not difficult.

Gao's algorithm consists in several steps that we will 
present below. We suppose that we are given parameters $k$, $\bc$ and
$\bg$ as above together with a codeword $c = \gamma_{k,\bc,\bg}(P)$
for an Ore polynomial $P$ of degree less than $k$. 
Let $w$ denote the ceiling of $\frac{n-k}2$ and 
let $e \in K^{n_1} \times \cdots \times K^{n_s}$ be 
a vector of rank-Hamming weight at most $w$. We set $m = c + e$.

\begin{example}[Thread example]
We shall illustrate each step of Gao's algorithm by the following 
thread example. As
in Example~\ref{eq:gamma}, we take $K = \FF_3(t)$ (equipped with 
$\theta = \id$ and $\partial = \frac d{dt}$), $k = 2$, $\bc = (0,1)$
and $\bg = ((1,t,t^2),(1,t,t^2))$. The generator matrix of
the corresponding RSG code is the matrix \eqref{eq:gamma}.
We will work with the following codeword:
$$c = \gamma_{k,\bc,\bg}\big(t^2 X + 1\big)
    = \big((1,\,t^2{+}t,\,2t^3{+}t^2), (t^2{+}1,\,t^3{+}t^2{+}t,\,t^4{+}2t^3{+}t^2)\big)$$
and the following error $e = \big((1,\,t^3,\,2t^3), (t{+}1,\,0,\,t^4{+}t^3)\big)$
which has rank-Hamming weight $2$. The corresponding received message is:
$$m = \big(
(2,\,t^3{+}t^2{+}t,\,t^3{+}t^2), (t^2{+}t{+}2,\,t^3{+}t^2{+}t,\,2t^4{+}t^2)\big).$$
\end{example}

\paragraph{Step 0: Annihilator.}

We compute the Ore polynomial~$L$ defined in \eqref{eq:defL}.

\noindent
If a fast multiplication algorithm of Ore polynomials is available
(which is notably the case when $\partial = 0$~\cite{pushwach,carleb}), 
this computation can be done efficiently by a divide-and-conquer
algorithm~\cite{carleb}.

We underline that this computation is independant of the received
message $m$ and then has to be done just once when the RSG code is 
set up.

\begin{example}
In our thread example, we have $L(X) = X^6 - X^3$ as shown by
Example~\ref{ex:L}.
\end{example}

\paragraph{Step 1: Interpolation.}

We compute a Ore polynomial~$\tilde P$ of degree less than $n$
such that $\gamma_{\bc,\bg}(P) = m$.

\noindent
This can be done for example by inverting the $K$-linear map 
$\gamma_{n,\bc,\bg}$, which is known to be a bijection by 
Lemma~\ref{lem:llcmaij}. Alternatively, $\tilde P$ can be
computed by solving a (noncommutative) Chinese remainder
problem. This latter approach is faster when an efficient
multiplication algorithm of Ore polynomials is available.

\begin{example}
In our thread example, we find:
$$\tilde P =
(2t^4{+}t^2) X^4 + (2t^4{+}t^3{+}2t) X^3 +
(2t^4{+}t^3{+}2t^2)X^2 + (t^3{+}t^2{+}2t) X + 2.$$
\end{example}

\begin{remark}
In general, it is possible that denominators appear and
that the degrees in $t$ get bigger than the maximal degree
in $t$ in $c$ and $m$. However, this growing always stays
under control.
\end{remark}

\paragraph{Step 2: Partial \rgcd.}

We compute a relation of the form
$U \tilde P + V L = R$
for Ore polynomials $U$, $V$ and $R$ with $\deg U \leq w$ and
$\deg R < w+k$.

\noindent
This relation can be computed by applying the extended Euclidean 
algorithm with the input $(\tilde P, L)$ and stopping it the first
time the remainder $R$ has degree less than $w+k$.

\begin{remark}
Using the theory of resultants and subresultants~\cite{li},
one can carry out this computation by controlling the degrees
in $t$ of all intermediate polynomials.
\end{remark}

\begin{example}
In our thread example, after one step in Euclidean algorithm,
we obtain:
$$\begin{array}{l}
  \big((2t{+}1)X^2 + tX\big) \cdot \tilde P 
+ (2t^5{+}t^4{+}t^3{+}2t^2) \cdot L \smallskip \\
\hspace{1cm}
= (2t^3{+}t^2) X^3 + (t^3{+}2t^2{+}1) X^2 + (2t^2{+}2t{+}2)X
\end{array}$$
so that we can take:
\begin{align*}
U & = (2t{+}1)X^2 + tX, \quad
V   = 2t^5{+}t^4{+}t^3{+}2t^2 \\
\text{and} \quad
R & = (2t^3{+}t^2) X^3 + (t^3{+}2t^2{+}1) X^2 + (2t^2{+}2t{+}2)X.
\end{align*}
\end{example}

The next proposition is the key result on which Gao's algorithm is
based.

\begin{proposition}
\label{prop:gao}
With the above notations, we have the relation $R = UP$
where $P$ is the Ore polynomial we used to construct the codeword 
$c$.
\end{proposition}

\paragraph{Step 3: Left Euclidean division.}
We compute the quotient $Q$ in the \emph{left} Euclidean division 
of $R$ by $U$. 

\noindent
By Proposition \ref{prop:gao}, $c = \gamma_{k,\bc,\bg}(Q)$ and we 
have decoded the message $m$.

\begin{example}
In our thread example, the left Euclidean division of $R$ by $U$
reads $R = U \cdot (1 + t^2 X)$; we have then reconstructed the
Ore polynomial $P$ we started with.
\end{example}

\end{document}